\documentclass{amsart}

\begin{document}

\newcommand{\nc}{\newcommand}

\nc{\pr}{\noindent{\em Proof. }}
\nc{\g}{\mathfrak g}
\renewcommand{\k}{\mathfrak k}
\nc{\A}{\mathcal A}
\nc{\F}{\mathcal F}
\renewcommand{\H}{\mathfrak H}

\newtheorem{theorem}{Theorem}{}
\newtheorem{lemma}[theorem]{Lemma}{}
\newtheorem{corollary}[theorem]{Corollary}{}
\newtheorem{conjecture}[theorem]{Conjecture}{}
\newtheorem{proposition}[theorem]{Proposition}{}
\newtheorem{axiom}{Axiom}{}
\newtheorem{remark}[theorem]{Remark}{}
\newtheorem{example}{Example}{}
\newtheorem{exercise}{Exercise}{}
\newtheorem{definition}{Definition}{}

\renewcommand{\theproposition}{\thesection.\arabic{proposition}}

\renewcommand{\thelemma}{\thesection.\arabic{lemma}}

\renewcommand{\thecorollary}{\thesection.\arabic{corollary}}

\renewcommand{\theremark}{}

\renewcommand{\thedefinition}{\arabic{definition}}

\renewcommand{\thetheorem}{\thesection.\arabic{theorem}}

\title{The mass gap problem for the Yang--Mills Field}

\author[A. Sevostyanov]{A. Sevostyanov}

\address{Institute of Pure and Applied Mathematics,
University of Aberdeen, Aberdeen AB24 3UE, United Kingdom, 
and 
\newline Max Planck Institute for Mathematics,
Vivatsgasse 7,
53111 Bonn,
Germany}
\email{a.sevastyanov@abdn.ac.uk \\}

\thanks{\noindent{\em 2000 Mathematics Subject Classification}  70S15, 81T13, 81T20 \\
{\em Key words and phrases.} Yang--Mills field}

\begin{abstract}
We consider the reduced Hamiltonian of the Yang--Mills field on $\mathbb{R}^4$ equipped with a Lorentzian metric.
We show that the secondary quantized principal term $H_0$ of the Taylor expansion of this Hamiltonian at the lowest energy point has a mass gap if and only if zero is not a point of the spectrum of the auxiliary self--adjoint operator ${\rm curl}=*d$ defined on the space of one-forms $\omega$ on $\mathbb{R}^3$ satisfying the condition ${\rm div}~ \omega=*d*\omega=0$, where $*$ is the Hodge star operator associated to a metric on $\mathbb{R}^3$ and $d$ is the exterior differential. In this case the classical lowest energy point of the reduced configuration space is a non--degenerate critical point of the potential energy term of the reduced Hamiltonian of the Yang--Mills field, in the sense of Palais.  
\end{abstract}

\maketitle

\pagestyle{myheadings}

\markboth{A. SEVOSTYANOV}{COMPLETE INTEGRABILITY AND THE Yang--Mills FIELD}

%%%%%%%%%%%%%%%%%%%%%%%%%%%%%%%%%%%%%%%%%%%%%%%%%%%%%%%%%%%%%%%%%%%%%%%%%%%%%%%%%%%%%%%%%%%%%%%%%%%%%%%%%%%%%%%%%%%%%%%%%%%%%%%%%%%%%%%%%%%%%%%%%%%%%%%%%%%%

\section*{Introduction}

\renewcommand{\theequation}{\arabic{equation}}

\setcounter{equation}{0}

The main objective of this paper is the study of the classical and of the quantum low energy behavior of the reduced Hamiltonian of the Yang--Mills field on $\mathbb{R}^4$ equipped with a non--degenerate Lorentzian metric $g_{\mu\nu}$, $\mu,\nu=0,1,2,3$ of signature $(+,-,-,-)$ and associated to a compact semisimple Lie group $K$ with Lie algebra $\k$. 

It is well known that the Yang--Mills field is an example of Hamiltonian systems with first class constraints in Dirac's terminology (see Sect. \ref{YMH} and \cite{Dir}), and hence its effective reduced phase space is obtained by Hamiltonian reduction, namely by the reduction of the cotangent bundle to the space of connections $\mathcal{D}$ on $\mathbb{R}^3$ with respect to the action induced by the action of the gauge group $\mathcal{K}$ on $\mathcal{D}$. The Hamiltonian $H$ of the Yang--Mills field is invariant under this action and gives rise to a reduced Hamiltonian $H_{red}$ on the reduced phase space (see Sect. \ref{YMPh}, \ref{redcoord}). $H$ is expressed in terms of a natural Riemannian metric $<\cdot,\cdot>$ on $\mathcal{D}$ induced by the Killing form of $\k$ and by the metric $g_{\mu\nu}$. This metric is invariant under the gauge action. Thus $\mathcal{D}/\mathcal{K}$ is equipped with a metric as well. For the sake of the Hamiltonian formulation we choose synchronous coordinates on $\mathbb{R}^4$, so $g_{00}=1$, $g_{0i}=0$, and assume that $-g_{ij}=k_{ij}$ is a positive definite metric on $\mathbb{R}^3$ independent of $t=x^0$. 

The Hamiltonian $H_{red}$ is quadratic in canonical momenta (its kinetic energy term is the square of the canonical momentum with respect to the metric $<\cdot,\cdot>$) and its behavior at low energies is essentially guided by the leading term $H_0$ of its Taylor expansion at the point of the reduced space corresponding to the gauge orbit of the trivial connection. $H_0$ is quadratic in both canonical coordinates and momenta and can be reduced to a standard form using the standard simultaneous diagonalization of the two quadratic forms defining the kinetic and the potential energy terms in the Hamiltonian $H_0$. This amounts to diagonalization of the self--adjoint operator ${\rm curl}=*d$ on the space of one-forms $\omega$ on $\mathbb{R}^3$ with values in $\k$ and satisfying the condition ${\rm div}~ \omega=*d*\omega=0$, where $*$ is the Hodge star operator associated to the metric $k_{ij}$ on $\mathbb{R}^3$ and $d$ is the exterior differential. 

After that the standard secondary quantization procedure can be applied to $H_0$, and we show that the spectrum of the quantized Hamiltonian $H_0$ has a mass gap $m$, i.e. the infimum of the spectrum of the secondary quantized Hamiltonian $H_0$ is $m$, if and only if zero does not belong to the spectrum of the operator ${\rm curl}$. In this case $m$ is the distance from zero to the spectrum of ${\rm curl}$. 

The spectrum of ${\rm curl}$ is closely related to the spectrum of the Hodge-Laplace operators on one-forms and on functions on $\mathbb{R}^3$. In particular, if the spectra of both of these operators do not contain zero then the spectrum of the quantized Hamiltonian $H_0$ has a mass gap. 

In the case of the Laplace operator defined on functions a review of the relevant results can be found e.g. in \cite{D2, Ur}. For instance, if the sectional curvature of the Levi--Civita connection of $k_{ij}$ is asymptotically bounded at infinity by a negative number then the corresponding Laplace operator has a gap in the spectrum separating in from zero (see \cite{D1}). 

A mass gap can also be achieved even if zero belongs to the spectrum of ${\rm curl}$ by modifying the reduced phase space. Namely, the Hamiltonians $H_{red}$ and $H_0$ should be restricted to a certain subspace of the reduced phase space which is defined in  such a way that the spectrum of the quantization of the restriction of $H_0$ to this subspace has a mass gap. The modification amounts to a certain cut-off procedure which presumably should naturally appear as a renormalization procedure in the course of a proper quantization of $H_{red}$. 

If the quantized Hamiltonian $H_0$ has a mass gap then at the classical level the Hamiltonian $H_{red}$ acquires an interesting property: the gauge orbit of the trivial connection becomes a non--degenerate critical point of the potential energy term of $H_{red}$ in the sense of Palais (see \cite{Pal}). As one can observe all field theories with a mass gap have this property. 

The Morse--Palais lemma asserts now that there is a local coordinate chart in $\mathcal{D}/\mathcal{K}$ containing the gauge orbit of the trivial connection in which the potential energy term of $H_{red}$ is equal to its second differential regarded as a quadratic form in coordinates on $\mathcal{D}$. Thus in the corresponding chart of the reduced phase space      
$H_{red}$ is reduced to the sum of the square of the momentum with respect to the Riemannian metric on $\mathcal{D}/\mathcal{K}$ and of a quadratic potential term. On this chart one can consider $H_{red}$ as a deformation of $H_0$ by means of a deformation of a constant metric on $\mathcal{D}/\mathcal{K}$ in terms of which $H_0$ is expressed, to the metric $<\cdot,\cdot>$.   

\vskip 0.3cm
\noindent
{\bf Acknowledgments.}

The author is grateful to Max-Planck-Institute for Mathematics, Bonn, where this work was completed.

%%%%%%%%%%%%%%%%%%%%%%%%%%%%%%%%%%%%%%%%%%%%%%%%%%%%%%%%%%%%%%%%%%%%%%%%%%%%%%%%%%%%%%%%%%%%%

\renewcommand{\theequation}{\thesection.\arabic{equation}}

%%%%%%%%%%%%%%%%%%%%%%%%%%%%%%%%%%%%%%%%%%%%%%%%%%%%%%%%%%%%%%%%%%%%%%%%%%%%%%%%%%%%%%%%%%%%%%%%%%%%%%%%%%%%%%%%%%%%%%%%%%%%%%%%%%%%%%%%%%%%%%%%%%%%%%%%%%%%%%

%%%%%%%%%%%%%%%%%%%%%%%%%%%%%%%%%%%%%%%%%%%%%%%%%%%%%%%%%%%%%%%%%%%%%%%%%%%%%%%%%%%%%%%%%%%

\section{The Yang--Mills field in the Hamiltonian formulation}\label{YMH}

\setcounter{equation}{0}
\setcounter{theorem}{0}
In this section we recall the Lagrangian and the Hamiltonian formalism for the Yang--Mills field in presence of external gravity (see e.g. \cite{Kif}). The canonical variables and the Hamiltonian will be obtained via the Legendre transform starting from the Lagrangian formulation. The general framework for the Hamiltonian approach in field theory is due to Dirac (see \cite{Dir,Dir1}). We follow the approach developed in \cite{FS} for the Yang--Mills field.

Let $K$ be a compact semisimple Lie group, $\k$ its Lie algebra and $\g$ the
complexification of $\k$. We denote by $(\cdot ,\cdot )$ the Killing form of $\g$.
Recall that the restriction of this form to $\k$ is non--degenerate and negatively defined.
We shall consider the affine space of smooth
connections in the trivial $K$-bundle, associated to
the adjoint representation of $K$, over the four--dimensional space--time $M=\mathbb{R}^4$ with the standard coordinates $(x^0,x^1,x^2,x^3)$.
Fixing the standard trivialization of the trivial $K$-bundle and the
trivial connection as an origin in the affine space of connections we can identify this space
with the space $\Omega^1(M,\k)$ of $\k$-valued 1-forms on $M$.
Let $\A\in \Omega^1(M,\k)$ be such a connection which is also called sometimes the Yang--Mills field.
Denote by $\F$ the curvature 2-form of this connection, $\F=d\A + \frac{1}{2} [\A\wedge \A]$.
Here as usual we denote by $[\A\wedge \A]$ the
operation which takes the
exterior product of $\k$-valued 1-forms and the commutator of their values in
$\k$.  We assume that $M$ is equipped with a non--degenerate metric $g_{\mu\nu}$, $\mu,\nu=0,1,2,3$ of signature $(+,-,-,-)$. We assume that the spatial part $g_{ij}$, $i,j=1,2,3$ of the metric is negatively defined and denote $k_{ij}=-g_{ij}$, so that $k_{ij}$ is positively defined. Denote $g=|{\rm det}~g_{\mu\nu}|$ and $k=|{\rm det}~k_{ij}|$. Note that 
$$
g^{00}=\frac{{\rm det}~k_{ij}}{{\rm det}~g_{\mu\nu}}=\frac{k}{g}> 0
$$ 
as both ${\rm det}~k_{ij}$ and ${\rm det}~g_{\mu\nu}$ are negative.

We shall consider the Yang--Mills action functional $YM$ which is defined by
the formula
\begin{equation}\label{YM}
YM (\A) = \frac 12 \int_{M} (\F \wedge , *\F),
\end{equation}
where $*$ stands for the Hodge star operation associated to the
metric $g_{\mu\nu}$, and we evaluate the Killing form on the
values of $\F$ and $*\F$ and also take their exterior product.

Next we pass from the Lagrangian to the Hamiltonian
formulation for the Yang--Mills field. To this end one should use the modified
action $YM'$,

\begin{equation}\label{L'}
YM'= \int_{M} (\left(d\A + \frac{1}{2} [\A\wedge \A]- \frac 12 \F\right) \wedge , *\F),
\end{equation}
where $\A$ and $\F$ should be regarded as independent variables. The
equations of motion obtained from the action functional $YM'$ are equivalent
to those derived from action (\ref{YM}). Indeed, the equation for $\F$
following from (\ref{L'}) is just the definition of the curvature, and the
other equation becomes the equation derived from action (\ref{YM}) after expressing $\F$
in terms of $\A$.

We denote by $A$ the ``three--dimensional Euclidean part'' of $\A$,
$A=A_kdx^k$, where $A_k=\A_k$ for
$k=1,2,3$. We also introduce the ``electric'' field $E$ and the ``magnetic''
field $B$ associated to $\F$ as follows:
$$
\begin{array}{l}
E^k=\F^{k0}\sqrt{g}, \\
\\
B=*^3F,~ F=d_3A + \frac{1}{2} [A\wedge A],
\end{array}
$$
where $*^3$ is the Hodge star operation with respect to the metric $k_{ij}$ on $\mathbb{R}^3$ and $d_3$ is the exterior differential for exterior forms on $\mathbb{R}^3$ regarded as the subspace of $M$ with coordinates $(x^1,x^2,x^3)$. Thus $F$ is the ``three-dimensional'' spatial part of $\F$.

Using this notation action (\ref{L'}) can be rewritten, after integration by parts and up to the integral of a divergence, in
the following form:
\begin{eqnarray}
\qquad YM'=\int_{M} ( -(\partial_0 A_i,\sqrt{g}\F^{i0})-\frac{1}{2}(\F_{i0},\F^{i0}\sqrt{g})+\frac{1}{4}(\F_{ij},\F^{ij}\sqrt{g})-   \label{L''} \\
 -(\A_0,\partial_i(\sqrt{g}\F^{i0})+[A_i,\sqrt{g}\F^{i0}]))d^4x, \nonumber
\end{eqnarray}
where, as usual, $\partial_\mu=\frac{\partial}{\partial x^\mu}$.

Denote $C=C(A,E)=\partial_i(E^{i})+[A_i,E^{i}]$, and
introduce an orthonormal basis $T_a,~a=1,\ldots ,{\rm dim}~\k $ in $\k$ with respect to the
Killing form, $(T_a,T_b)=-\delta_{ab}$, and the components of $A$, $E$, $\A_0$ and $C$ associated to this basis,
$A_k=A_k^{a}T_a,~E^{k}=
E^{k,a}T_a,~\A_0=\A_0^{a}T_a,~C=C^{a}T_a$. After a straightforward and tedious calculation action (\ref{L''}) can be rewritten in terms of these components as follows
\begin{equation}\label{YM''}
 YM'=\int_{M}\left( E^{i,a}\partial_0 A_i^a-h(A,E)+\A_0^aC^a\right) d^4x,
\end{equation}
where 
$$
h(A,E)=\frac{1}{\sqrt{g^{00}}}\left(\frac{1}{2}\frac{1}{\sqrt{k}}k_{ij}E^{i,a}E^{j,a}+\frac{1}{4} F_{ij}^aF_{kl}^ak^{ik}k^{jl}\sqrt{k}\right)+g^{0j}F_{ij}^aE^{i,a},
$$
and 
$$
k^{ij}=-g^{ij}+\frac{g^{i0}g^{j0}}{g^{00}}
$$
is the inverse matrix to $k_{ij}$.
From formula (\ref{YM''}) it is clear that $A_k^{a}$ and $E^{k,a}$ are
canonical conjugate coordinates and momenta for the Yang--Mills field,
$h(A,E)$ is the Hamiltonian density, $\A_0^{a}$ are Lagrange multipliers and
$C^a=0$ are constrains imposed on the canonical variables.

The equations obtained from action (\ref{YM}) become Hamiltonian with respect to the canonical Poisson
structure
\begin{equation}\label{pois}
  \{E^{k,a}(x),A_l^b(y)\}=\delta^k_l\delta^{ab}\delta(x-y),
\end{equation}
and all the other Poisson brackets of the components of $E$ and $A$ vanish.
One can also check that
\begin{equation}\label{momentbrack}
\{C^a(x),C^b(y)\}=\sum_c t^{abc}C^c(x)\delta(x-y),
\end{equation}
where $t^{abc}$ are the structure constants of Lie algebra $\k$ with respect
to the basis $T_a$, $[T_a,T_b]=\sum_c t^{abc}T_c$, and that
\begin{equation}\label{haminv}
\{H(A,E), C^a(x)\}=0,
\end{equation}
where 
$$
H(A,E)=\int_{\mathbb{R}^3}h(A,E)d^3x
$$
is the Hamiltonian. This means that the Yang--Mills field is a generalized Hamiltonian system
with first class constrains according to Dirac's classification (see \cite{Dir}).

%%%%%%%%%%%%%%%%%%%%%%%%%%%%%%%%%%%%%%%%%%%%%%%%%%%%%%%%%%%%%%%%%%%%%%%%%%%%%%%%%%%%%%%%%%%

\section{The structure of the phase space of the Yang--Mills field}\label{YMPh}

\setcounter{equation}{0}
\setcounter{theorem}{0}

In this section we collect some facts on the Poisson geometry of the phase
space of the Yang--Mills field and related gauge actions. These results are well known (see \cite{BV, FS, FU, Sing, Sol, W}).

Consider the affine space  of smooth connections
in the trivial $K$-bundle over ${\mathbb R}^{3}$ associated to
the adjoint representation of $K$. As in Section
\ref{YMH} we fix the standard trivialization of this bundle and the
trivial connection as an origin in the affine space of connections and identify this space
with the space $\Omega^1(\mathbb{R}^{3},\k)$ of $\k$-valued 1-forms on ${\mathbb
R}^{3}$. Let $\mathcal{D}$ be the space of $W_2^1$--Sobolev connections (see \cite{FU}, \S 3 or \cite{W}, Appendix B). $\mathcal{D}$ is the space of $\k$-valued one--forms on  $\mathbb{R}^{3}$ with components in $W_2^1(\mathbb{R}^3)\otimes \k$, where in the definition of the Sobolev space the components are differentiated with the help of the covariant derivative of the Levi-Civita connection corresponding to the metric $k_{ij}$.

The tangent space $T_A\mathcal{D}$ to $\mathcal{D}$ at point $A\in \mathcal{D}$ is isomorphic to $\mathcal{D}$.
We define a scalar product on this space by
\begin{equation}\label{prod}
<\omega_1,\omega_2>=-\int_{\mathbb{R}^{3}}(\omega_1\wedge,*^3\omega_2),
\end{equation}
where $\omega_{1,2} \in T_A\mathcal{D}$, $*^3$ stands for the Hodge star operation associated to the metric $k_{ij}$ on  $\mathbb{R}^{3}$, and we evaluate the Killing form on the
values of $\omega_1$ and $*\omega_2$ and also take their exterior product. Thus the space $\mathcal{D}$ is equipped with a Riemannian metric.   

Let $T^*_A\mathcal{D}$ be the space of $W_2^1$-Sobolev $\k$-valued contravariant densities on $\mathbb{R}^{3}$ of weight one with respect to the metric $k_{ij}$. Elements of this space are of the form $\sqrt{k}V$, where $V$ is $\k$-valued vector field on $\mathbb{R}^{3}$ with components in $W_2^1(\mathbb{R}^3)\otimes \k$. 

There is a natural paring between this space and the space $T_A\mathcal{D}$ which we denote by the same symbol as the scalar product on $T_A\mathcal{D}$,
\begin{equation}\label{par}
<E,A>=\int_ME^{i,a}A_i^ad^3x,~E=E^{i,a}T_a\frac{\partial}{\partial x^i}\in T^*_A\mathcal{D}, A=A_i^a T_a d x^i\in T_A\mathcal{D}.
\end{equation}

Using this paring one can define an induced scalar product on $T^*_A\mathcal{D}$ which we denote again by the same symbol as the scalar product on $T_A\mathcal{D}$,
$$
<Z,W>=\int_M\frac{1}{\sqrt{k}}k_{ij}Z^{i,a}W^{j,a}d^3x,~Z=Z^{i,a}T_a\frac{\partial}{\partial x^i},W=W^{i,a}T_a\frac{\partial}{\partial x^i}\in T^*_A\mathcal{D}.
$$

Pairing (\ref{par}) gives rise to an embedding of the space $T^*_A\mathcal{D}$ into the space dual to $T_A\mathcal{D}$. Therefore the bundle $T^*\mathcal{D}$ over $\mathcal{D}$ the fiber of which at point $A\in \mathcal{D}$ is $T^*_A\mathcal{D}$ naturally becomes a subbundle of the cotangent bundle to $\mathcal{D}$.

Using the last imbedding the bundle $T^*\mathcal{D}$ can be
equipped with the natural structure of a Poisson manifold induced by
the canonical symplectic structure of the cotangent bundle to $\mathcal{D}$.

Assume from now on that the coordinates $x^\mu$ are synchronous for the metric $g_{\mu\nu}$, i.e. $g_{00}=g^{00}=1$, $g^{0i}=g_{0i}=0$, and hence $g=k$. We shall also always suppose that $k_{ij}$ is a positive definite metric on $\mathbb{R}^3$ independent of $t=x^0$. In this case Poisson structure (\ref{pois}) has a natural geometric interpretation in terms of the Poisson structure of $T^*\mathcal{D}$.

Indeed, for each $A$ the ``electric'' field $E$ belongs to the space of $\k$-valued contravariant densities on $\mathbb{R}^{3}$ of weight one with respect to the metric $k_{ij}$ which contains $T^*_A\mathcal{D}$ as a subspace. The Poisson structure (\ref{pois}) on the space $T^*\mathcal{D}$ can be identified with the natural Poisson structure on it induced by the canonical symplectic structure of the cotangent bundle to $\mathcal{D}$.

Now let us discuss the meaning of the constrains. First of all we note that
the constrains $C(A,E)$ infinitesimally generate the gauge action on
the phase space $T^*\mathcal{D}$. More precisely, consider the group
$\mathcal{K}$ of $K$--valued $W_2^2$--Sobolev maps $h:\mathbb{R}^3\to K$ satisfying the condition $\lim_{x\to \infty}h(x)=I$, where $I$ is the identity element of $K$ (for the definition of $\mathcal{K}$ see \cite{W}, Appendix A, \cite{FU}, \S 3 or \cite{Sol} for the case of a non-compact base). $\mathcal{K}$ is a Lie group called the gauge group. 

The gauge group $\mathcal{K}$ continuously acts on the space of connections $\mathcal{D}$ (see e.g. \cite{W}, Lemma A.6, \cite{Sol}, Theorem 1)
by
\begin{equation}\label{Gaugeact}
\begin{array}{l}
\mathcal{K}\times \mathcal{D} \rightarrow \mathcal{D},\\
\\
h\times A \mapsto h\circ A= -dhh^{-1}+hAh^{-1},
\end{array}
\end{equation}
where we denote $hAh^{-1}=\mathrm{Ad}h(A)$.

The action (\ref{Gaugeact}) of $\mathcal{K}$ on the space of connections
$\mathcal{D}$ induces an action
\begin{equation}\label{Gaugeactph}
\begin{array}{l}
\mathcal{K}\times T^*\mathcal{D}\rightarrow T^*\mathcal{D}, \\
\\
h\times (A,E) \mapsto (-dhh^{-1}+hAh^{-1}, hEh^{-1}),
\end{array}
\end{equation}
where we write $hEh^{-1}=\mathrm{Ad}h(E)$. This action gives rise
to an action of the Lie algebra of the gauge
group $\mathcal{K}$ on $T^*\mathcal{D}$ by vector fields. If $X$ is an element of the Lie algebra of $\mathcal{K}$
then the corresponding vector field $V_X(A,E)$ is given by
\begin{equation}\label{gaugeactph}
V_X(A,E)=(-dX+[X,A],[X,E]),~E\in T^*_A\mathcal{D}, A\in \mathcal{D}\simeq T_A\mathcal{D}.
\end{equation}
Here we, of course, identify
$T_{(A,E)}T^*\mathcal{D}\simeq T_A\mathcal{D}\times T_A^*\mathcal{D}$.

The gauge action (\ref{Gaugeact}) is free, so that the quotient $\mathcal{D}/\mathcal{K}$ is a smooth
manifold (see \cite{Sing}). Indeed, if $X$ belongs to the Lie algebra of the stabilizer of a point $A$ then $dX+[A,X]=0$, and by the invariance of the Killing form we have
$$
d(X,X)=(dX+[A,X],X)+(X,dX+[A,X])=0,
$$
i.e. $(X,X)$ is constant, and hence $\lim_{x\to \infty}X(x)\neq 0$, i.e. $X$ is not in the Lie algebra of $\mathcal{K}$.

The action (\ref{gaugeactph}) is generated by the constraint $C(A,E)$ in the sense that if $X$ is an element of the Lie algebra of $\mathcal{K}$ and $(A,E)\in T^*\mathcal{D}$ then we have
$$
\{\int_{\mathbb{R}^3}(C(A,E),X)d^3x,A(x)\}=-dX(x)+[X(x),A(x)],
$$
and
$$
\{\int_{\mathbb{R}^3}(C(A,E),X)d^3x,E(x)\}=[X(x),E(x)].
$$

Using the language of Poisson geometry and taking into account formula (\ref{momentbrack})
for the Poisson brackets of the constraints one can say that
$\mathcal{K}\times T^*\mathcal{D}\rightarrow T^*\mathcal{D}$ is a Hamiltonian
group action, and the map
\begin{equation}\label{YMmoment}
\begin{array}{l}
\mu(A,E)=C(A,E)
\end{array}
\end{equation}
is a moment map for this action. In particular, action (\ref{Gaugeactph})
preserves the Poisson structure of $T^*\mathcal{D}$.

We note that action (\ref{Gaugeact}) also preserves the Riemannian structure of the configuration
space $\mathcal{D}$. This follows from the fact that the Killing form on $\k$ is invariant with respect to the adjoint action of $K$.

The properties of the phase space of the Yang--Mills field
and of the gauge action discussed above are formulated in the following proposition.
\begin{proposition}\label{YMprop}
Let $\mathcal{D}$ be the space of $W_2^1$--Sobolev $K$--connections on
$\mathbb{R}^{3}$, $\mathcal{K}$ the corresponding group of $W_2^2$--Sobolev gauge transformations vanishing at infinity. Then

(i) The space $\mathcal{D}$ is an infinite dimensional Riemannian
manifold equipped with metric (\ref{prod}).
The bundle $T^*\mathcal{D}$ can be
equipped with the natural structure of a Poisson manifold induced by
the canonical symplectic structure of the cotangent bundle to $\mathcal{D}$;

(ii) The gauge action $\mathcal{K}\times \mathcal{D}\rightarrow \mathcal{D}$
preserves Riemannian metric (\ref{prod}) and
gives rise to a Hamiltonian group action
$\mathcal{K}\times T^*\mathcal{D}\rightarrow T^*\mathcal{D}$ with the moment map
$$
\begin{array}{l}
\mu(A,E)=C(A,E),~(A,E)\in T^*\mathcal{D};
\end{array}
$$

(iii) The action of the gauge group $\mathcal{K}$ on the spaces $\mathcal{D}$ and $T^*\mathcal{D}$
is free, and the reduced phase space $\mu^{-1}(0)/\mathcal{K}$ is a smooth manifold.
\end{proposition}

Finally we make a few remarks on the structure of the Hamiltonian of the
Yang--Mills field.

In terms of the above defined scalar products, and under the assumption that the coordinates $x^\mu$ are synchronous for the metric $g_{\mu\nu}$, the Hamiltonian $H(A,E)$ takes a very simple form,
\begin{equation}\label{hamilt}
H(A,E)=\frac 12 (<E,E>+<B,B>). 
\end{equation}

Since the Hamiltonian $H(A,E)$ of the Yang--Mills field is invariant under gauge action (\ref{Gaugeactph}) (this fact can be checked directly and also follows from formula (\ref{haminv})) the generalized Hamiltonian
dynamics described by this Hamiltonian together with the constrains $C(A,E)=0$
is equivalent to the usual one on the
reduced phase space $\mu^{-1}(0)/\mathcal{K}$ (see \cite{A,Dir}).

The Hamiltonian (\ref{hamilt}) itself has a very standard structure; $H(A,E)$ is
equal to the sum of a half of the square of the canonical momentum, $\frac 12
<E,E>$, and of a potential $U(A)$, $U(A)=\frac 12 <B,B>$. The potential $U(A)$ is, in
turn, equal to a half of the square of the vector field $B\in
\Gamma(T\mathcal{D})$. By definition the vector field $B$ is invariant with respect
to the gauge action of $\mathcal{K}$, $B(g\circ A)=gB(A)g^{-1}$.

Let $\Omega^n(\mathbb{R}^{3},\k)$ be the space of
$\k$-valued differential $n$--forms on $\mathbb{R}^{3}$. We recall that the covariant derivative
$d_A:\Omega^n(\mathbb{R}^{3},\k)\rightarrow \Omega^{n+1}(\mathbb{R}^{3},\k)$ associated to a connection $A\in \Omega^1(\mathbb{R}^{3},\k)$
is defined by $d_A\omega =d_3\omega + [A\wedge \omega]$, and the operator $d_A^*:\Omega^{n+1}(\mathbb{R}^{3},\k)\rightarrow \Omega^{n}(\mathbb{R}^{3},\k)$
formally adjoint to $d_A$ with respect to the scalar product on $\Omega^*(\mathbb{R}^{3},\k)$ given by formula (\ref{prod}) is
equal to $d_A^*=(-1)^{3(n+1)}*^3d_A*^3$. We denote by ${\rm div}_A$ the part of this operator acting
from $\Omega^1(\mathbb{R}^{3},\k)$ to $\Omega^{0}(\mathbb{R}^{3},\k)$, with
the opposite sign,
$$
{\rm div}_A=*^3d_A*^3:\Omega^1(\mathbb{R}^{3},\k)\rightarrow
\Omega^{0}(\mathbb{R}^{3},\k).
$$

Note that under the isomorphism 
$$
T_A^*\mathcal{D}\simeq T_A\mathcal{D},~E\mapsto E'
$$
induced by pairing (\ref{par}) the operator ${\rm div}_A$ corresponds to $C(A,\cdot)$. More precisely, if $E=\sqrt{k}E^i\frac{\partial}{\partial x^i} \in T_A^*\mathcal{D}$ then the corresponding element of $T_A\mathcal{D}$ is $E'=E^ik_{ij}dx^j$, and $C(A,E)=\sqrt{k}~{\rm div}_AE'$.

The value of the vector field $B$ at each point $A\in \mathcal{D}$
belongs to the kernel of the operator ${\rm div}_A$,
${\rm div}_AB(A)=0$ for all $A\in \mathcal{D}$. Indeed, from the Bianci identity
$d_AF=0$, the definition
of $B=*^3F$ and the formula $*^3*^3=id$ it follows that
$$
{\rm div}_A~B=*^3d_A*^3*^3F=*^3d_A~F=0.
$$

Now we summarize the properties of the Hamiltonian of the Yang--Mills field.
\begin{proposition}\label{YMhamiltprop}
(i) The generalized Hamiltonian system on the Poisson manifold $T^*\mathcal{D}$ with the
Hamiltonian $H(A,E)$,
$H(A,E)=\frac 12 (<E,E>+<B,B>)$, $B=*^3F$, $F=d_3A+\frac{1}{2}[A\wedge A]$, and the constrains
$C(A,E)=0$
describes the Yang--Mills dynamics on $T^*\mathcal{D}$.

(ii) The Hamiltonian $H(A,E)$ is invariant under the gauge action
$\mathcal{K}\times T^*\mathcal{D}\rightarrow T^*\mathcal{D}$ and the generalized Hamiltonian
dynamics described by this Hamiltonian together with the constrains $C(A,E)=0$
is equivalent to the usual one on the reduced phase space
$\mu^{-1}(0)/\mathcal{K}$.

(iii) The vector field $B$ is invariant with respect
to the gauge action of $\mathcal{K}$, $B(g\circ A)=gB(A)g^{-1}$.
The value of this field at each point $A\in \mathcal{D}$
belongs to the kernel of the operator ${\rm div}_A$,
${\rm div}_AB(A)=0$ for all $A\in \mathcal{D}$.
\end{proposition}

%%%%%%%%%%%%%%%%%%%%%%%%%%%%%%%%%%%%%%%%%%%%%%%%%%%%%%%%%%%%%%%%%%%%%%%%%%%%%%%%%%%%%%%%%%%%%%%%

\section{The structure of the reduced phase space: a model case}\label{redcoord}

\setcounter{equation}{0}
\setcounter{theorem}{0}

In Propositions \ref{YMprop} and \ref{YMhamiltprop} we formulated all properties of the
Yang--Mills field which are important for our further consideration. In this section we
study an arbitrary Hamiltonian system satisfying these properties. For simplicity, in this section we only consider the finite-dimensional case.

First we consider a phase space equipped with a Lie group action of the type
described in Proposition \ref{YMprop}. Actually
the Riemannian metric introduced in that proposition is only important for the definition
of the Hamiltonian of the Yang--Mills field. This metric is not relevant to
Poisson geometry. We used this metric in the description of the phase space in order to avoid
analytic difficulties arising in the infinite-dimensional case. Now let us
forget about the metric for a moment and discuss the geometry of the reduced
space.

The Poisson structure described in Proposition \ref{YMprop} is an example of
the canonical Poisson structure on the cotangent bundle, and the group action
on this bundle is induced by a group action on the base manifold.
Thus we start with a manifold $\mathcal{M}$ and a Lie group
$G$ freely acting on $\mathcal{M}$. The canonical symplectic structure on $T^*\mathcal{M}$
can be defined as follows (see \cite{A}).

Denote by $\pi:T^*\mathcal{M}\rightarrow \mathcal{M}$ the
canonical projection, and define a 1-form $\theta$ on $T^*\mathcal{M}$ by
$\theta (v)=p(d\pi v)$,
where $p\in T_x^*\mathcal{M}$ and $v\in T_{(x,p)}(T^*\mathcal{M})$. Then the canonical
symplectic form on $T^*\mathcal{M}$ is equal to $d\theta$.

Recall that the induced Lie group action
$G\times T^*\mathcal{M}\rightarrow T^*\mathcal{M}$ is a Hamiltonian group action
with a moment map $\mu:T^*\mathcal{M}\rightarrow \g^*$, where $\g^*$ is the
dual space to the Lie algebra $\g$ of $G$. The moment map $\mu$ is
uniquely determined by the formula (see \cite{Per}, Theorem 1.5.2)
\begin{equation}\label{mom}
(\mu(x,p),X)=\theta (\widehat {\widehat X})(x,p)=p(\widehat X(x)),
\end{equation}
where $\widehat{X}$ is the vector field on
$\mathcal{M}$ generated by an arbitrary element $X\in \g$, $\widehat {\widehat X}$ is the
induced vector field on $T^*\mathcal{M}$ and $(,)$ stands for the
canonical paring between $\g$ and $\g^*$.

Formula (\ref{mom}) implies that for
any $x\in \mathcal{M}$ the map $\mu(x,p)$ is linear in $p$. We denote this linear map by $m(x)$,
$m(x):T_x^*\mathcal{M}\rightarrow \g^*$,
\begin{equation}\label{m}
m(x)p=\mu(x,p).
\end{equation}

Next, following \cite{A}, Appendix 5, with some modifications of the proofs suitable for
our purposes, we describe the structure of the reduced space
$\mu^{-1}(0)/{G}$. We start with a simple lemma.
\begin{lemma}\label{l1}
The annihilator $T_x\mathcal{O}^\perp\in T_x^*\mathcal{M}$ of the tangent
space $T_x\mathcal{O}$ to the $G$-orbit $\mathcal{O}\subset \mathcal{M}$ at point $x$ is
isomorphic to ${\rm Ker}~m(x)$, $T_x\mathcal{O}^\perp={\rm Ker}~m(x)$.
\end{lemma}
\pr
First we note that the space $T_x\mathcal{O}^\perp$ is spanned by the
differentials of $G$-invariant functions on $\mathcal{M}$. But from the
definitions of the moment map and of the Poisson structure on
$T^*\mathcal{M}$ we have
\begin{equation}\label{to}
L_{\widehat{X}}f(x)=\{(X,\mu),f\}(x)=(X,m(x)df(x)),
\end{equation}
where $\widehat{X}$ is the vector field on $\mathcal{M}$ generated by
element $X\in \g$, $f\in C^{\infty}(\mathcal{M})$, and $(,)$ stands for the
canonical paring between $\g$ and $\g^*$.

Formula (\ref{to}) implies that $f$ is $G$-invariant if and only if
$df(x)\in {\rm Ker}~m(x)$. This completes the proof.

\qed

\begin{proposition}\label{redstruct}
The action of the group $G$ on $T^*\mathcal{M}$
is free, and the
reduced phase space $\mu^{-1}(0)/{G}$ is a smooth manifold. Moreover, we have
an isomorphism of symplectic manifolds, $\mu^{-1}(0)/{G}\simeq
T^*(\mathcal{M}/{G})$, where $T^*(\mathcal{M}/{G})$ is equipped with the
canonical symplectic structure.
Under this isomorphism
$T_{\mathcal{O}_x}^*(\mathcal{M}/{G})\simeq T_x\mathcal{O}_x^\perp$, where
$\mathcal{O}_x$ is the $G$-orbit of $x$.
\end{proposition}

\pr
Let $\mathcal{O}_x$ be the  $G$-orbit of point $x\in \mathcal{M}$ and
$\pi:\mathcal{M}\rightarrow \mathcal{M}/{G}$ the canonical projection,
$\pi(x)=\mathcal{O}_x$. Denote by $\Xi$ the foliation of the space $\mathcal{M}$ by the
subspaces $T_x\mathcal{O}^\perp$. Since the foliation $\Xi$ is $G$-invariant
and
${\rm Ker}~d\pi|_{T_x\mathcal{M}}=T_x\mathcal{O}_x$ we can identify the
subspace $T_x\mathcal{O}_x^\perp$ with the space
$T_{\mathcal{O}_x}^*(\mathcal{M}/{G})$ by means of the dual map to
the differential of the projection
$\pi$. But the definition of the moment map $\mu$ and Lemma \ref{l1}
imply that
$\mu^{-1}(0)=\{(x,p)\in T^*\mathcal{M}:p \in T_x\mathcal{O}_x^\perp\}$. Therefore
the quotient $\mu^{-1}(0)/{G}$ is diffeomorphic to
$T^*(\mathcal{M}/{G})$, the diffeomorphism being induced by the canonical
projection $\pi$.

From the definitions of
the Poisson structures on $T^*(\mathcal{M}/{G})$ and on the reduced space
$\mu^{-1}(0)/{G}$ it follows that the diffeomorphism
$\mu^{-1}(0)/{G}\simeq T^*(\mathcal{M}/{G})$ is actually an isomorphism of symplectic
manifolds.

\qed

Using the last proposition one can easily describe the space $\Gamma T^*(\mathcal{M}/{G})$ of
covector fields on $\mathcal{M}/{G}$.
\begin{corollary}\label{vectred}
The space $\Gamma T^*(\mathcal{M}/{G})$ is isomorphic to the space of
$G$-invariant sections $V\in \Gamma T^*\mathcal{M}$ such that $V(x)\in
T_x\mathcal{O}_x^\perp$ for any $x\in \mathcal{M}$. Such covector fields
will be called horizontal $G$-invariant covector fields on $\mathcal{M}$. We
denote this space by $\Gamma_G^\perp T^*\mathcal{M}$,
$\Gamma_G^\perp T^*\mathcal{M}\simeq \Gamma T^*(\mathcal{M}/{G})$.
\end{corollary}

Now we discuss the class of Hamiltonians on $T^*\mathcal{M}$ we are
interested in. First, recalling Proposition \ref{YMprop} we equip the manifold
$\mathcal{M}$ with a Riemannian metric $<,>$ and
assume that the action of $G$ on $\mathcal{M}$  preserves
this metric. Using this metric we can establish an isomorphism of
$G$-manifolds, $T\mathcal{M}\simeq T^*\mathcal{M}$.
We shall always identify the tangent and the cotangent bundle of
$\mathcal{M}$ and the spaces of vector and covector fields on $\mathcal{M}$
by means of this isomorphism.
The tangent bundle $T\mathcal{M}$
will be regarded as a symplectic manifold with the induced symplectic structure.
Under the identification $T\mathcal{M}\simeq T^*\mathcal{M}$ the subspace
$T_x \mathcal{O}^\perp\subset T_x^*\mathcal{M}$ is isomorphic to the
orthogonal complement of the tangent space $T_x\mathcal{O}$ in
$T_x\mathcal{M}$. Note also that since
$T_{\mathcal{O}_x}^*(\mathcal{M}/{G})\simeq T_x\mathcal{O}_x^\perp$,
and the metric on $\mathcal{M}$ is $G$-invariant,
$T_{\mathcal{O}_x}^*(\mathcal{M}/{G})$ has a scalar product induced
from $T_x\mathcal{O}_x^\perp$, i.e. $\mathcal{M}/{G}$ naturally
becomes a Riemannian manifold. We shall also identify
$T^*(\mathcal{M}/{G})\simeq T(\mathcal{M}/{G})$ by means of the metric.
Denote by $\Gamma_G^\perp T\mathcal{M}$ the space of $G$-invariant horizontal vector fields on
$\mathcal{M}$, i.e. vector fields $V(x)$ satisfying the condition $V(x)\in T_x\mathcal{O}_x^\perp$ for any $x\in \mathcal{M}$. By Corollary \ref{vectred} we have an isomorphism,
$\Gamma_G^\perp T\mathcal{M}\simeq \Gamma T(\mathcal{M}/{G})$.

On the symplectic manifold $T\mathcal{M}$ we define a Hamiltonian $H$ of the
type described in Proposition \ref{YMhamiltprop}. In order to do that we fix
a $G$-invariant horizontal vector field $V$ on $\mathcal{M}$. Then we put
$$
H(x,p)=\frac 12 (<p,p>+<V(x),V(x)>),~p\in T_x\mathcal{M}.
$$
This Hamiltonian is obviously $G$-invariant and gives rise to a reduced Hamiltonian
$H_{red}$ on the reduced space $\mu^{-1}(0)/{G}\simeq T^*(\mathcal{M}/{G})$.
Since by Corollary \ref{vectred} $V$ can be regarded as a (co)vector field on $\mathcal{M}/{G}$
we have
\begin{equation}\label{hamiltred}
H_{red}(\mathcal{O}_x,p_\perp)=\frac 12
(<p_\perp,p_\perp>+<V(x),V(x)>),~p_\perp \in T_x\mathcal{O}_x^\perp\simeq
T_{\mathcal{O}_x}^*(\mathcal{M}/{G}).
\end{equation}

%%%%%%%%%%%%%%%%%%%%%%%%%%%%%%%%%%%%%%%%%%%%%%%%%%%%%%%%%%%%%%%%%%%%%%%%%%%%%%%%%%%%%%%%%%%%%%%%%%%%%%%%%%%%%%%%

\section{Low energy behavior of the Yang--Mills Hamiltonian  on
the reduced phase space and its quantization}\label{YMcoord}

In this section we describe a low energy approximation $H_0$ to the Hamiltonian of the Yang--Mills field on the reduced phase space (see Proposition \ref{YMhamiltprop}) and obtain a necessarily and sufficient condition under which the spectrum of the secondary quantized Hamiltonian $H_0$ has a mass gap. This condition implies that the gauge orbit of the trivial connection is a non--degenerate critical point for the potential $U(A)$ of the Hamiltonian of the Yang--Mills field on the space $\mathcal{D}/\mathcal{K}$.    

First observe that according to Proposition \ref{YMhamiltprop} the vector field $B(A)$ on the space $\mathcal{D}$ is $\mathcal{K}$--invariant and horizontal. Therefore from the last observation made in the previous section we infer that Hamiltonian (\ref{hamilt}) gives rise to the reduced Hamiltonian 
\begin{equation}\label{Hamiltred}
H_{red}(\mathcal{O}_A,E_\perp)=\frac 12
(<E_\perp,E_\perp>+<B(A),B(A)>),~C(A,E_\perp)=0 
\end{equation}
on the reduced space $\mu^{-1}(0)/\mathcal{K}$.

As we observed above under the isomorphism $T_A^*\mathcal{D}\simeq T_A\mathcal{D}$, $E\mapsto E'$ induced by pairing (\ref{par}) the operator ${\rm div}_A$ corresponds to $C(A,\cdot)$, and hence the space $T_{\mathcal{O}_A}\mathcal{D}/\mathcal{K}$ is isomorphic to the kernel of the operator ${\rm div}_A$ in $T_A\mathcal{D}$. The metric (\ref{prod}) induces a Riemannian metric on $\mathcal{D}/\mathcal{K}$ which we denote by the same symbol. 

Consider the scalar product on the space $\Omega^*_c(\mathbb{R}^3,\g)$ of all compactly supported $\g$-valued forms on $\mathbb{R}^3$ given by the formula
\begin{equation}\label{prodc}
<\omega_1,\omega_2>=-\int_{\mathbb{R}^{3}}(\omega_1\wedge,*\overline{\omega}_2),~
\omega_{1,2} \in \Omega^*(\mathbb{R}^{3},\g),
\end{equation}
where $\overline{\omega}_2$ is the complex conjugate of $\omega_2$ with respect to the
complex structure induced by the decomposition $\g=\k\stackrel{\cdot}{+}i\k$.
Note that $\k\subset \g$ is a real subspace with respect to this complex
structure.

Let $\mathcal{H}^1$ be the completion of the space $\Omega^1_c(\mathbb{R}^3,\g)$ of compactly supported $\g$-valued 1-forms on $\mathbb{R}^3$ with respect to scalar product (\ref{prodc}).
We also denote by $\mathcal{H}^0$ the completion of the space
$\Omega^0_c(\mathbb{R}^3,\g)$ with respect to scalar product (\ref{prodc}).
Then, clearly, the operator $\mathrm{div}_A$ is naturally extended to a linear operator
${\rm div}_A:\Omega^1_c(\mathbb{R}^3,\g)\rightarrow
\Omega^0_c(\mathbb{R}^3,\g)$ and the closure of this operator is a well-defined closed operator
${\rm div}_A:\mathcal{H}^1\rightarrow \mathcal{H}^0$. Denote
by $P_A:\mathcal{H}^1\rightarrow {\rm Ker}~{\rm div}_A$
the orthogonal projection operator onto the kernel of this operator. Note that both
$\mathrm{div}_A$ and $P_A$ preserve the natural real structure on
$\mathcal{H}^1$, i.e. they are real operators.

Locally the reduced configuration space $\mathcal{D}/\mathcal{K}$ can be described as follows. For every point $A\in \mathcal{D}$ there exists an open neighborhood such that each $\mathcal{K}$--orbit locally intersects ${\rm Ker}~{\rm div}_A\cap \mathcal{D}$ at a unique point (see, e.g., \cite{Sing}). Therefore ${\rm Ker}~{\rm div}_A\cap \mathcal{D}$ plays the role a local model (coordinate chart) for $\mathcal{D}/\mathcal{K}$ in a neighborhood of the orbit $\mathcal{O}_A$ of $A$.

We are primarily interested in the behavior of $H_{red}(\mathcal{O}_A,E_\perp)$ in a neighborhood of the orbit $\mathcal{O}_0$ of the trivial connection. Let $\mathcal{C}={\rm Ker}~{\rm div}_0\mid_{\mathcal{D}}$ be the coordinate chart for this neighborhood. For brevity we shall also write ${\rm div}_0={\rm div}$.  $T^*\mathcal{C}=\{(A,E), A\in \mathcal {C}, E\in T^*_0(\mathcal{D}), C(0,E)=0\}\simeq \{(A,E'), A,E'\in \mathcal {C}\}$ is a coordinate chart for $T^*\mathcal{D}/\mathcal{K}$ near the orbit $\mathcal{O}_0$ of the trivial connection.

Observe that $P_A$ is the orthogonal projection operator onto ${\rm Ker}~{\rm div}_A$, and ${\rm Ker}~{\rm div}_A\mid_{\mathcal{D}}\simeq T_{\mathcal{O}_A}\mathcal{D}/\mathcal{K}$ as linear spaces equipped with scalar product induced by (\ref{prod}). Therefore in an open neighborhood of the orbit $\mathcal{O}_0$ and in terms of the local chart $T^*\mathcal{C}$ for this neighborhood the Hamiltonian $H_{red}$ takes the form
\begin{equation}\label{redH}
H_{red}(A,E')=\frac 12
(<P_A E',P_A E'>+<B(A),B(A)>),A\in W\cap \mathcal{C},E'\in \mathcal{C}, 
\end{equation}
where $W\subset \mathcal{C}$ is an open subset containing $A=0$ (see \cite{BV}, Sect. 4 for more details on the metric in terms of the coordinate chart $T^*\mathcal{C}$). Since $B(A)=*dA+*[A\wedge A]$, the first term of the Taylor expansion at $A=0$ of Hamiltonian (\ref{redH}) is
\begin{equation}\label{YMhamred}
H_0(A,E')=\frac 12
(<E',E'>+<*d A,*d A>),A,E'\in \mathcal {C}.
\end{equation}

To study this Hamiltonian we have to simultaneously diagonalize the quadratic forms $<E',E'>$ and $<*d A,*d A>$. This is achieved by observing that ${\rm curl}=*d$ gives rise to a self-adjoint operator ${\rm curl}:{\rm Ker}~{\rm div}\to {\rm Ker}~{\rm div}$ with the domain ${\rm Ker}~{\rm div}\cap \mathcal{D}_\mathbb{C}$, where $\mathcal{D}_\mathbb{C}$ is the complexification of $\mathcal{D}$ (see e.g. \cite{S3}, Section 3.6).  

Let
$$
\Psi: \mathrm{Ker}~\mathrm{div}\rightarrow 
\int_{\mathbb{R}}H(\lambda)d\mu(\lambda)=\mathcal{H}
$$
be the corresponding generalized Fourier
transform, where $\int_{\mathbb{R}}H(\lambda)d\mu(\lambda)$ is a direct Hilbert space integral and $\mu$ is a measure on $\mathbb{R}$. Note that integration in the formula above is actually taken over the support ${\rm supp}~\mu$ of $\mu$ which coincides with the spectrum $\sigma ({\rm curl})$ of the operator ${\rm curl}=*d$ on $\mathrm{Ker}~\mathrm{div}$. 
Note that $\Psi{\rm curl}\Psi^{-1}=M(\lambda)$ is the operator of multiplication by $\lambda$.

Let $e_i(\lambda)$, $i\in \mathbb{N}$ be an orthonormal basis in the Hilbert space $H(\lambda)$, so every element $f(\lambda)\in H(\lambda)$ is uniquely represented as the sum of a converging series $f(\lambda)=\sum_{i=1}^\infty f_i(\lambda)e_i(\lambda)$, $f_i(\lambda)\in \mathbb{C}$.

Now using the generalized Fourier transform $\Psi$ one can introduce other canonical coordinates on $T^*\mathcal{C}$,
\begin{eqnarray}\label{YMcoordredg1}
q_{\lambda ,i}(A)=\Psi(A)_{i}(\lambda),\\
p_{\lambda ,i}(E)=\Psi(E')_{i}(\lambda), \label{YMcoordredp1}
\end{eqnarray}
where we denote by $\Psi(\cdot )_{i}(\lambda)$ the components of the
Fourier transform $\Psi$ with respect to the basis $e_i(\lambda)$, $i\in \mathbb{N}$.
They have the following canonical Poisson brackets
$$
\{p_{\lambda ,i},q_{\lambda' ,j}\}=\delta_{ij}\delta_\mu(\lambda-\lambda'),
$$
where $\delta_\mu(\lambda-\lambda')$ is the delta function associated to the measure $\mu$.

In terms of coordinates (\ref{YMcoordredg1}), (\ref{YMcoordredp1}) 
Hamiltonian (\ref{YMhamred}) can be rewritten in the following form
$$
H_0=\int_{{\rm supp}~\mu}
\sum_{i=1}^\infty (|p_{\lambda,i}|^2+|\lambda |^2|g_{\lambda
,i}|^2)d\mu(\lambda).
$$

Introducing other new coordinates
$$
a_{\lambda ,j}=\frac{1}{\sqrt{2|\lambda|}}(|\lambda|q_{\lambda ,j}+ip_{\lambda ,j}),~~a^*_{\lambda ,j}=\frac{1}{\sqrt{2|\lambda|}}(|\lambda|\overline{q}_{\lambda ,j}-i\overline{p}_{\lambda ,j})
$$
with Poisson brackets
$$
\{a_{\lambda,k},a^*_{\lambda',j}\}=
i\delta_{kj}\delta_\mu(\lambda-\lambda')
$$
we can rewrite $H_0$ in the form
\begin{equation}\label{H0}
H_0=\int_{{\rm supp}~\mu}
\sum_{i=1}^\infty |\lambda|a^*_{\lambda,i}a_{\lambda
,i}d\mu(\lambda).
\end{equation}

After applying the secondary quantization procedure  $a^*_{\lambda,i},~a_{\lambda
,i}$, $\lambda \in {\rm supp}~\mu$ become the creation and the annihilation operators in the Fock space generated by the ground state $|0>$ satisfying the conditions
$$
a_{\lambda
,i}|0>=0~{\rm for~ all}~\lambda \in {\rm supp}~\mu,~i\in \mathbb{N}.
$$
They obey the commutation relations
$$
[a_{\lambda,i},a^*_{\lambda',j}]=
\delta_{ij}\delta_\mu(\lambda-\lambda').
$$

The generalized eigenvectors of the quantized Hamiltonian $H_0$ are of the form 
$$
a^*_{\lambda_1,i_1}\ldots a^*_{\lambda_n,i_n}|0>,~n\in \mathbb{N}, \lambda_i \in \sigma ({\rm curl})={\rm supp}~\mu,
$$
and they have eigenvalues 
\begin{equation}\label{eigenv}
|\lambda_1|+\ldots +|\lambda_n|.
\end{equation}

However, the spectrum $\sigma ({\rm curl})$ depends on the metric $g_{\mu\nu}$. From (\ref{eigenv}) it follows that if this metric is such that $\sigma ({\rm curl})$ does not contain point zero then the spectrum of the secondary quantized Hamiltonian $H_0$ has a mass gap equal to the distance $m$ from the point zero to the set $\sigma ({\rm curl})$. Note that various types of the spectrum of this Hamiltonian may occur depending on $\sigma ({\rm curl})$, e.g. in addition to the absolutely continuous spectrum a point spectrum corresponding to bound states or a singular spectrum may occur.

One can study the spectrum of ${\rm curl}$ by considering
the selfadjoint operator $T$,
\begin{equation}\label{s0}
T=\left(
\begin{array}{cc}
{\rm curl} & -d \\
  {\rm div} & 0
\end{array}
\right),
\end{equation}
acting in the space $\mathcal H^1\stackrel{\cdot}{+}\mathcal H^0$ with the
natural domain
$$
\mathfrak D (T)=\{(\omega,u)\in\mathcal H^1\stackrel{\cdot}{+}\mathcal H^0:{\rm curl}~\omega,
du\in\mathcal{H}^1,{\rm div}~\omega\in \mathcal{H}^0\}
$$
(see \cite{BS1,BS2}). One has 
\begin{equation}\label{s02}
T^2=\left(
\begin{array}{cc}
\widehat{\triangle} & 0 \\
  0 & \triangle
\end{array}
\right),
\end{equation}
where $\triangle=-{\rm div}~d$ is the usual Laplace operator and $\widehat{\triangle}={\rm curl}~{\rm curl}-d_3{\rm div}$ is the Laplace-Hodge operator on 
$\mathcal H^1$. ${\rm Ker}~{\rm div}\subset \mathcal H^1$ is an invariant subspace for $T$ and $T\mid_{{\rm Ker}~{\rm div}}={\rm curl}$.
Therefore $(\sigma ({\rm curl}))^2\subset \sigma(T^2)$. Thus if $0\not\in \sigma(T^2)$ then $0\not\in \sigma ({\rm curl})$, and the problem of the mass gap is reduced to the study of the spectra of the operators $\triangle$ and $\widehat{\triangle}$. In particular, if $0\not\in \sigma(\triangle)\cup \sigma(\widehat{\triangle})=\sigma(T^2)$ then the quantized Hamiltonian $H_0$ has a mass gap. 

If $0\in \sigma ({\rm curl})$ then one can still get a mass gap $m$ by quantizing the restriction of $H_0$ to the subspace $T^*\mathcal{C}_m$,
$$
\mathcal{C}_m=\{x\in \mathcal{C}: \Psi(x)\in \mathcal{H}_m\},
$$
where
$$
\mathcal{H}_m=\{f\in \mathcal{H}: f(\lambda)=0~{\rm for}~|\lambda|< m\},
$$
and $m>0$ is a parameter. Formula similar to (\ref{H0}) will still hold, where the integration is performed over ${\rm supp}~\mu\setminus (-m,m)$. 

At the classical level the condition $0\not\in \sigma ({\rm curl})$ implies that the operator $M(\lambda)$ is a bounded isomorphism of its domain
$$
\mathfrak{D}(M(\lambda))=\{f(\lambda)\in \mathcal H: \int_{{\rm supp}~\mu}\lambda^2|f(\lambda)|^2d\mu(\lambda)<\infty \}
$$
with the norm given by
$$
\Vert f(\lambda)\Vert+\Vert\lambda f(\lambda)\Vert,
$$
where $\Vert\cdot \Vert$ is the norm in $\mathcal{H}$,
onto $\mathcal{H}$. Indeed, the bounded inverse operator is given by the multiplication by $\frac1\lambda$ since this function is bounded measurable on ${\rm supp}~\mu$.
Recalling that ${\rm Ker}~{\rm div}\cap \mathcal{D}_\mathbb{C}$ is the domain of the real operator ${\rm curl}$ isomorphic to $\mathfrak{D}(M(\lambda))$ under $\Psi$ and that $\mathcal{H}\simeq \mathcal{H}^*$, since $\mathcal{H}$ is a Hilbert space, we deduce that the bilinear form $<{\rm curl}~ \omega,{\rm curl}~ \omega'>$, $\omega,\omega'\in \mathcal{C}$ equal to the second differential of $\frac12<B(A),B(A)>$ at $A=0\in \mathcal{C}$ gives rise to an isomorphism 
$$
\mathcal{C}\simeq \mathcal{C}^*, \omega\mapsto <{\rm curl}~ \omega,{\rm curl}~ \cdot>.
$$

Therefore $A=0\in \mathcal{C}$ is a non-degenerate critical point of $U(A)=\frac12<B(A),B(A)>$. Observe that, as a function of $A$, $U(A)$ is a continuous polynomial function on $\mathcal{C}$. Therefore it is differentiable infinitely many times and according to the Morse-Palais lemma (see \cite{Pal}) there is a neighborhood $V$ of $A=0\in \mathcal{C}$ and a $C^\infty$--diffeomerphism $\Phi:V\to V$ such that, if we denote $Q=\Phi(A)$, then 
$$
U(\Phi^{-1}(Q))=\frac12<{\rm curl}~ Q,{\rm curl}~ Q>.
$$
Thus if we write $P=(d\Phi^*)^{-1}E$, $E\in T^*_0(\mathcal{D}),~C(0,E)=0$ and denote by $G(\cdot,\cdot)$ the metric on $T\mathcal{C}\simeq T^*\mathcal{C}$ in the corresponding variables $P'\in \mathcal{C}$ then
\begin{equation}\label{finH}
H_{red}(Q,P)=\frac 12
(G(P',P')+<{\rm curl}~Q,{\rm curl}~Q>),Q\in V, P'\in \mathcal{C}. 
\end{equation}

We conclude that $H_{red}$ is a member of a family of Hamiltonians parametrized by Riemannian metrics on $\mathcal{C}$ and with the same potential term. In particular, when $G$ is replaced with the constant metric $<\cdot,\cdot>$ we get $H_0$.

We summarize the discussion above in the following theorem.
\begin{theorem}
The spectrum of the secondary quantized Hamiltonian $H_0$ has a mass gap $m>0$ if and only if $0$ is not a point of the spectrum of the operator  ${\rm curl}$. The gap $m$ is equal to the distance from zero to the spectrum of the operator  ${\rm curl}$. In this case $A=0\in \mathcal{C}$ is a non--degenerate critical point of the Hamiltonian $H_{red}$ and there is a local coordinate chart in $T\mathcal{D}/\mathcal{K}\simeq T^*\mathcal{D}/\mathcal{K}$ containing the gauge orbit of the trivial connection in which $H_{red}$ takes form (\ref{finH}).
\end{theorem}

\end{document}